\documentclass[aps,twocolumn,showpacs,superscriptaddress,pre,floatfix]{revtex4-2}
\usepackage{amsmath}
\usepackage{amssymb}

\usepackage{dcolumn}%

\usepackage{tikz}
\usepackage{mathtools}
\usepackage{hyperref}
\hypersetup{
    colorlinks,
    linkcolor={blue!80!black},
    citecolor={blue!80!black},
    urlcolor={blue!80!black}
}
\usepackage[normalem]{ulem}

\usepackage{todonotes}
\newcommand{\nn}{\text{NN}}
\newcommand{\nfmd}{n_\text{fmd}}

\begin{document}

\title{Accurate ground-state entropies from population Monte Carlo: The antiferromagnetic Ising model on the Shastry-Sutherland lattice}
\date{\today}

\author{Denis Gessert}%
\affiliation{Institut f\"ur Physik, Technische Universit\"at Chemnitz, 09107 Chemnitz, Germany}

\author{Wolfhard Janke}%
\affiliation{Institut f\"{u}r Theoretische Physik, Universit\"at Leipzig, IPF 231101, 04081 Leipzig, 
Germany}
  
\author{Martin Weigel}%
\affiliation{Institut f\"ur Physik, Technische Universit\"at Chemnitz, 09107 Chemnitz, Germany}
  
\begin{abstract}
    We demonstrate how the population annealing Monte Carlo simulation method can be used to compute the ground-state entropy to high precision. This approach is tested for the classical antiferromagnetic Ising model on the Shastry-Sutherland lattice with diagonal couplings chosen twice as strong as the nearest-neighbor interactions. %
    In the absence of an external magnetic field, the model is known to be disordered at all temperatures, with a macroscopically degenerate ground state. 
    We find the ground-state entropy to be $0.458\,777\,78(10)$ per site, which is in good agreement with the estimate of a recent study using the corner transfer matrix renormalization group method. The sampled ground-state configurations are analyzed by considering the bond configurations, the local-energy configurations, and the spin-spin correlations. While all observations are consistent with the disordered nature of the ground state, the numerical data suggest a power-law decay of the size distribution of clusters of sites with local energies different from the ground-state level.
\end{abstract}

\maketitle
\section{Introduction}
Geometric frustration in spin systems often leads to rich behavior, including partial order, reentrance, residual entropy, and exotic phases~\cite{Diep2013}. Among such systems, spin models on the Shastry-Sutherland lattice have attracted considerable attention over the last decades~\cite{Shastry1981,Nomura2023,Trinh2018,Yadav2025,Meng2008,Chang2009,Dublenych2012,Moliner2009,Grechnev2013,Huang2012,Farkasovsky2019,Yadav2025,Shastry2026}. %
For the Heisenberg model, for a wide range of parameters the quantum-mechanical ground state can be calculated exactly~\cite{Shastry1981}, and there are close experimental realizations such as SrCu$_2$(BO$_3$)$_2$~\cite{Nomura2023} and TmB$_4$~\cite{Trinh2018}.
Experimentally, multiple fractional magnetization plateaus are observed in Shastry-Sutherland compounds as the external field strength is varied~\cite{Trinh2018,Nomura2023,Yadav2025}. Even the simplest models such as the classical Ising~\cite{Meng2008,Chang2009,Dublenych2012} or Heisenberg~\cite{Moliner2009,Grechnev2013} antiferromagnets (AFM) on the Shastry-Sutherland lattice produce a fractional magnetization plateau, however, only at $1/3$ of the saturation magnetization~\cite{Grechnev2013,Meng2008,Chang2009,Dublenych2012} instead of showing a cascade of values as observed experimentally. Theoretical considerations indicate that in the Ising AFM a cascade can occur when introducing long-range dipole-dipole interactions~\cite{Huang2012}, third- and fourth-neighbor interactions~\cite{Farkasovsky2019}, or lattice distortions~\cite{Yadav2025} into the model.

In the absence of an external field, the zero-temperature classical Ising AFM on the Shastry-Sutherland lattice shows Néel order for weak second-neighbor interactions, whereas for stronger diagonal couplings it exhibits a macroscopically degenerate AFM dimer state~\cite{Shastry1981,Meng2008,Dublenych2012}.
At the limiting interaction strength that separates these two scenarios, the ground state is even more degenerate~\cite{Shastry1981,Meng2008,Shastry2026},
with the nature of it still not fully understood. This problem has recently been addressed by Shastry~\textit{et al.}~\cite{Shastry2026} by calculating, among other quantities, the residual entropy to high precision using the corner transfer matrix renormalization group (CTMRG) method.

In the present work, we take a somewhat different approach. In order to sample the ground states, we use the population annealing Monte Carlo method, which works well for systems with rugged free energy landscapes~\cite{Hukushima2003,Machta2010,Wang2015a,Weigel2021} even down to the ground states~\cite{Wang2015a}. We demonstrate how this technique can be used as a general tool for the estimation of the ground-state entropy. Our results are in good agreement with those of Ref.~\cite{Shastry2026}. Some exemplary ground-state configurations are presented in several different representations including one based on the local energy. Due to the strong frustration in this model, the local energies can take values different from the ground-state energy, and only averaging over the whole lattice yields the ground-state energy. While the ground states do not show any signs of long-range order, a cluster analysis of the local-energy representation numerically suggests a power-law cluster-size distribution of sites with local energies different from the ground-state energy. The associated power-law exponent is distinctly different from that of the random cluster model in two dimensions.

The rest of the paper is organized as follows. In Sec.~\ref{sec:modelObservable} the model and the relevant observables are introduced. The population annealing method is reviewed in Sec.~\ref{sec:methods}, and we show how it can be used to estimate the ground-state entropy. Key simulation details can be found in Sec.~\ref{sec:simDetails}. In Sec.~\ref{sec:results} we present our numerical results, and Sec.~\ref{sec:conclusionAndOutlook} contains our conclusion and an outlook to future work.

\section{Model and observables}\label{sec:modelObservable}
\begin{figure}
    \includegraphics{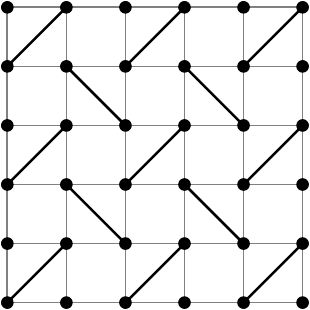}
    \caption{Shastry-Sutherland lattice of a $6\times 6$ square lattice.\label{fig:lattice}}
\end{figure}
We consider the antiferromagnetic Ising model on the Shastry-Sutherland lattice depicted in Fig.~\ref{fig:lattice} with Hamiltonian
\begin{equation}
    \mathcal{H} = \sum_{\langle ij\rangle} \sigma_i\sigma_j + 2\alpha \sum_{[ lm ]} \sigma_l\sigma_m \equiv \Sigma_\nn + 2\alpha \Sigma_\text{d},\label{eq:Hamiltonian}
\end{equation}
where $\sigma_i\in\{-1,1\}$ denote the $N=L \times L$ Ising spins, $\langle ij\rangle$ the nearest-neighbor (\nn) interactions (gray lines), and $[ lm ]$ the diagonal interactions (black thick lines) of the lattice. $\Sigma_\nn$ and $\Sigma_\text{d}$ are introduced as shorthand notations for the sums over \nn{} and diagonal interactions, respectively. Throughout, periodic boundary conditions (PBC) are assumed.

For $\alpha < 1$, the Néel state with energy $E^{(\text{Néel})} = (\alpha-2) N$ minimizes Eq.~\eqref{eq:Hamiltonian}, that is, the system is antiferromagnetically ordered at zero temperature with ground-state energy $E_\text{GS} = E^{(\text{Néel})}$. In this case, all \nn~interactions ($\Sigma_\nn^{(\text{Néel})}=-2N$) are satisfied, and all the diagonals are ferromagnetically oriented and therefore frustrated ($\Sigma_\text{d}^{(\text{Néel})}=N/2$). In contrast, for $\alpha > 1$, a dimer state is energetically more favorable where the diagonals are antiferromagnetic ($\Sigma_\text{d}^{(\text{dimer})}=-N/2$) at the expense of half of the \nn~interactions being frustrated ($\Sigma_\nn^{(\text{dimer})}=0$). Hence, $E_\text{GS}=E^{(\text{dimer})}=-\alpha N$. While the Néel state is two-fold degenerate, this dimer state is macroscopically degenerate, as each of the $N/2$ diagonal interactions has two energetically equivalent orientations, giving rise to $2^{N/2}$ ground states in total and whence a residual entropy of $\ln (2) / 2\approx 0.346\,574\dots$ per site. 

Here, we consider the case of $\alpha=1$, in which the Néel state and the dimer state both have the same energy, i.e., $E_\text{GS}=E^{(\text{Néel})}=E^{(\text{dimer})}=-N$. Thus, some diagonals are ferromagnetically ordered while others are not, giving rise to an even larger residual entropy than in the case $\alpha>1$~\cite{Shastry1981,Meng2008,Shastry2026}.
For $\alpha=1$ an alternative (but equivalent) way to write the Hamiltonian~\eqref{eq:Hamiltonian} is as a sum over all triplets of spins $[ijk]$ on a triangle, i.e., $\mathcal{H} = \sum_{[ijk]} (\sigma_i\sigma_j + \sigma_j \sigma_k + \sigma_k \sigma_i)$~\cite{Shastry2026}. In doing so each nearest-neighbor (diagonal) interaction is counted exactly once (twice), and hence this form is identical to Eq.~\eqref{eq:Hamiltonian}. Clearly, in each triangle at most two interactions can be satisfied, and the energy is minimized when each triangular plaquette is populated by two spins of one orientation with the third pointing in the other direction. Due to the shared diagonal interactions, in this case the residual entropy is not as simple to calculate as for $\alpha>1$. 
The authors of Ref.~\cite{Meng2008} estimate the zero-temperature entropy by the following simple argument: 
There are 16 possible spin configurations of four spins on a plaquette with a diagonal interaction, out of which only 10 are allowed in ground states. Using this fact, and that there are $N/2$ such plaquettes, they estimate the ground-state entropy as 
\begin{equation}
    \!\!\!\frac{S(T\!=\!0)}{N}\! \approx \!\frac 1 N\! \ln\!\!\left[ 2^N \!\!\left(\frac{10}{16}\right)^{\!\!\!\frac N 2}\!\right]\!\!=\! \frac 1 2 \ln{\!\left( \frac 5 2\right)} \!\!=\! 0.458\,145\!\dots\label{eq:entropyPlaquetteApprox}
\end{equation}
A recent corner transfer matrix renormalization group (CTMRG) study~\cite{Shastry2026} estimated it to be $S(T=0)/N=0.458\,778\dots$, which is close to but distinctly different from this value. The authors of Ref.~\cite{Shastry2026} also find that the fraction of ferromagnetically ordered diagonal interactions equals $\nfmd=0.203\,953\dots$.

Besides the entropy we study the following thermodynamic quantities: (i) The energy $E=\langle \mathcal{H} \rangle$, which takes values between $0$ and $-N$ with the former being the value at infinite temperature and the latter at zero temperature; (ii) the specific heat per site, $C_V = \beta^2 (\langle \mathcal{H}^2 \rangle - \langle \mathcal{H} \rangle^2) / N$, where $\beta = 1/T$ is the inverse temperature (setting $k_B=1$ to fix units); and (iii) the staggered magnetization $M_s = \sum_{x,y} (-1)^{x+y} \sigma_{x,y}$. The motivation to consider $M_s$ despite the lack of long-range order at zero temperature is that this would be the relevant order parameter for $\alpha<1$, and therefore may give some insight into the competition between \nn{} and diagonal interactions. (iv) We also consider the fraction of diagonal interactions whose spins are ferromagnetically aligned $\nfmd$ given by~\cite{Shastry2026}
    \begin{equation}
    \nfmd = \frac {1} {N} \sum_{[lm]} (1 + \sigma_l \sigma_m),
    \end{equation}
where as in Eq.~\eqref{eq:Hamiltonian} $\sum_{[lm]}$ refers to the diagonal interactions of Fig.~\ref{fig:lattice}. Since it is energetically preferred for diagonal interactions to be antiferromagnetically aligned $\nfmd$ also quantifies the fraction of diagonal interactions that are frustrated. Further, note that $\nfmd$ is trivially related to the diagonal energy term $\Sigma_\text{d}$ of the Hamiltonian energy $\mathcal{H}=\Sigma_\nn + 2\alpha \Sigma_\text{d}$ of Eq.~\eqref{eq:Hamiltonian} by $\nfmd=1/2+\Sigma_\text{d}/N$.

\section{Population annealing}\label{sec:methods}
Population annealing (PA)~\cite{Hukushima2003,Machta2010} is a more recent addition to the toolkit of advanced Monte Carlo (MC) methods, in which a population of $R$ replicas is collectively cooled down from easily accessible high temperatures towards low temperatures typically more difficult to sample. As the temperature is lowered, replicas are resampled such that phase space exploration of regions with higher Boltzmann weight is favored while giving up on replicas stuck in metastable states of low weight.

The algorithm works as follows. Initially, all $R$ replicas are drawn at random from the starting distribution. In the case of an Ising system, as is the case here, this is most conveniently done at $\beta_0=0$ by simple sampling. 
After initialization, one repeats the resampling and MC steps until the lowest temperature of interest $T_\text{f} = 1/\beta_\text{f}$ is reached. The resampling step consists of increasing the iteration counter $i$ by one (starting initially at $i=0$), and then resampling the replicas according to their Boltzmann weights at the new inverse temperature $\beta_i$. Hence, a replica with energy $E_k$ on average is copied 
\begin{equation}
    \hat \tau_i(E_k) = \frac {R}{R_{i-1}} \exp \left[-\left(\beta_i-\beta_{i-1}\right) E_k\right] / Q_i
\end{equation}
times, where
\begin{equation}
    Q_i = \frac{1}{R_{i-1}} \sum_{k=1}^{R_{i-1}} \exp \left(-(\beta_i-\beta_{i-1})E_k\right)
\end{equation}
is the necessary normalization to keep the population size (approximately) constant. Note that the number of copies refers to how many times a replica occurs in the new population at $\beta_i$, that is zero copies means a replica is discarded, one copy means that the replica survives once, two copies that it is copied once etc. We use the nearest-integer resampling-method~\cite{Wang2015}, but see Ref.~\cite{Gessert2023} for an overview of other possible resampling methods. Subsequently, one carries out $\theta_i$ MC updates for each replica, and observables are estimated via population averages. Note that the population size $R$, the annealing schedule $\{\beta_i\}$, and the sweep schedule $\{\theta_i\}$ are parameters that can be tuned for optimal PA performance, see Sec.~\ref{sec:simDetails} for the simulation details.

Clearly $Q_i$ is an estimator for $\langle e^{-(\beta_i-\beta_{i-1})E}\rangle_{\beta_{i-1}}$, which, as is easy to show, is equal to the ratio of the partition functions at $\beta_i$ and $\beta_{i-1}$, i.e., $\langle Q_i \rangle = Z(\beta_i) / Z(\beta_{i-1})$~\cite{Machta2010}.
Since $Z(\beta_0)$ is usually known (here, $2^N$), this in principle allows estimating $Z(\beta_i)$ at all simulation temperatures~$\beta_i$ which by standard techniques would only be known up to prefactors. Thus, a natural estimator for the free energy $-\beta F=\ln Z$ is given by~\cite{Machta2010} 
\begin{equation}
    -\beta_i F_i = \ln Z(\beta_0) + \sum_{k=1}^{i} \ln Q(\beta_{k-1},\beta_k).
\end{equation}
From this the entropy can easily be estimated at any simulation temperature via $S(\beta) = \beta E(\beta) - \beta F(\beta)$.

While $S(T)$ is close to the ground-state entropy for low temperatures, taking it as an estimate for $S(T=0)$ clearly will in general result in overestimating the true ground-state entropy. When the value of the ground-state energy $E_\text{GS}$ is known, one can measure the fraction of the population that has reached the ground state, $\rho_\text{GS}(\beta_i)$, and from this estimate the ground-state entropy via
\begin{equation}
    S(T=0) = S(\beta_i) + \ln[\rho_\text{GS}(\beta_i)] - [E(\beta_i)-E_\text{GS}] \beta_i \label{eq:pa_S_GS_estimator}
\end{equation}
at finite inverse temperatures $\beta_i$, for which the measured value of $\rho_\text{GS}(\beta_i)$ is larger than zero. 
To arrive at Eq.~\eqref{eq:pa_S_GS_estimator}, we first note that the zero-temperature entropy is the logarithm of the number of ground states, i.e., $S(T=0)= \ln \Omega(E_\text{GS})$.
Further, the probability of a system to be in the ground state (and hence, up to fluctuations, the fraction of replicas in the ground state) is given by
\begin{equation}
    \rho_\text{GS}(\beta_i) = \frac{1}{Z(\beta_i)} \Omega(E_\text{GS}) e^{-\beta_i E_\text{GS}}.\label{eq:probGroundState}
\end{equation}
Rearranging Eq.~\eqref{eq:probGroundState}, and using $S(T=0)= \ln \Omega(E_\text{GS})$ yields
\begin{equation}
    S(T=0) = \ln[Z(\beta_i)] + \ln[\rho_\text{GS}(\beta_i)] + E_\text{GS}\beta_i.\label{eq:proofEntropyAlmostDone}
\end{equation}
With the thermodynamic identity $\ln[Z(\beta_i)]=-\beta_i F(\beta_i) = S(\beta_i)-E(\beta_i)\beta_i$ and Eq.~\eqref{eq:proofEntropyAlmostDone} one arrives at Eq.~\eqref{eq:pa_S_GS_estimator}.

This approach can also be applied in cases where the ground-state energy is unknown by relying on an ``optimistic'' entropy estimate. By replacing the ground-state energy in Eq.~\eqref{eq:pa_S_GS_estimator} with the lowest sampled energy $E_{\min}$, the estimator gives correct results iff the ground-state energy was indeed found, and otherwise it estimates $\Omega(E_{\min})$. This is an upper bound of the actual ground-state entropy per site assuming that $\Omega(E)$ is a monotonically increasing function near $E_\text{GS}$. Note that this assumption usually holds, and is equivalent to requiring that the microcanonical temperature $\hat T = (\mathrm{d} [\ln \Omega(E)] / \mathrm{d} E)^{-1}$ is positive.
\section{Simulation details}\label{sec:simDetails}
\begin{figure}[t]
    \centering
    \includegraphics{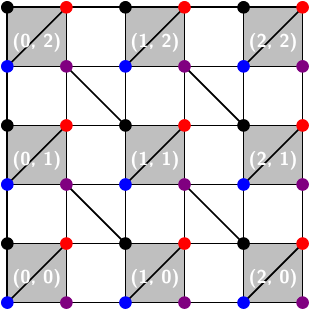}
    \caption{Domain decomposition of the Shastry-Sutherland lattice into four sublattices used for GPU updates. For better memory coalescence, each sublattice is stored separately. Spins on a plaquette highlighted in gray share the same index, with the respective two-dimensional indices indicated.\label{fig:latticeDomainDecomposition}}
\end{figure}

The code of our PA-GPU simulations is based on the open-source implementation of Ref.~\cite{Barash2017}. Besides the parallelism on replica level inherent to PA, Ref.~\cite{Barash2017} uses a checkerboard domain decomposition to simulate the $d=2$ Ising model. While this domain decomposition cannot be used directly due to the diagonal interactions, it is straightforward to find a suitable domain decomposition into four sublattices as illustrated in Fig.~\ref{fig:latticeDomainDecomposition}. For better memory coalescence spins of different sublattices are stored separately~\cite{Barash2017}. The gray squares in Fig.~\ref{fig:latticeDomainDecomposition} denote plaquettes of spins with the same indices (white text) in their respective arrays.

As in the original implementation of Ref.~\cite{Barash2017}, the nearest-integer resampling method~\cite{Wang2015} is used and adaptive temperature steps. Regarding the choice of the resampling method, we have shown in Ref.~\cite{Gessert2023} that nearest-integer resampling performs favorably compared to other resampling methods. For the adaptive temperature protocol we use a target histogram overlap between $87\%$ and $90\%$. In addition, we use an adaptive sweep protocol as discussed in Ref.~\cite{Gessert2024} requiring that the effective population size~\cite{Weigel2021} is $R_\text{eff}(E) \geq 0.95$.
The population size $R$ is always chosen as large as possible such that the entire population fits into the memory of the GPU used~\footnote{For this study, all simulation runs were carried out on Nvidia RTX A5000 GPUs with 24GB memory each (one GPU per simulation).} (see Table~\ref{tab:resultsEntropy} below).
All results were averaged over $M=20$ independent simulations each. Error estimates are obtained from calculating the standard error of the independent runs.

\section{Results}\label{sec:results}
\begin{figure}
\includegraphics{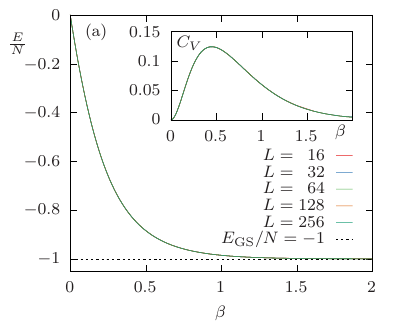}
\includegraphics{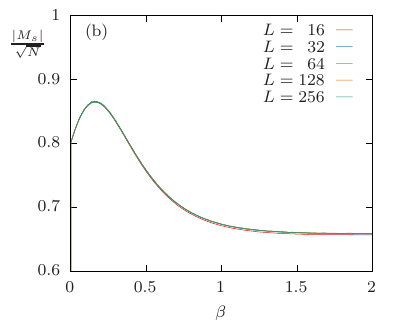}
\caption{(a) Energy per site $E/N$, and (b) (staggered) magnetization per square-root-volume $|M_s|/\sqrt{N}$ as function of inverse temperature $\beta$ for different system sizes $L$. In both cases, finite-size effects are negligible. The inset of panel (a) shows the specific heat per site~$C_V$. Error bars are smaller than the line width.\label{fig:ecm}}    
\end{figure}

\begin{figure}[h]
\includegraphics{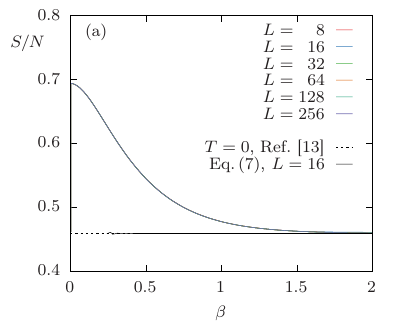}
\includegraphics{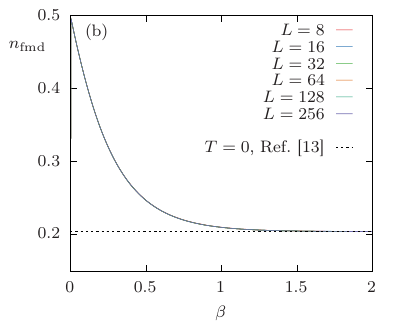}
\caption{(a) Entropy per site $S/N$, and (b) the fraction of ferromagnetically oriented diagonal interactions $\nfmd$ as function of inverse temperature $\beta$ for different system sizes $L$. In both cases, the zero-temperature results from Ref.~\cite{Shastry2026} are indicated as dotted lines, and as for the energy and magnetization in Fig.~\ref{fig:ecm} finite-size effects are negligible. In panel (a), the estimator for the residual entropy~\eqref{eq:pa_S_GS_estimator} evaluated for the $L=16$ simulation data is shown as a solid black line, and it is in good visual agreement with the value of Ref.~\cite{Shastry2026}. Error bars are smaller than the line width.\label{fig:nfmd_and_S}}
\end{figure}

To get an overview of the thermodynamic behavior for $\alpha=1$, we first consider the energy $E/N$, the specific heat $C_V$, and the absolute value of the staggered magnetization $|M_s|$ as a function of inverse temperature $\beta=1/T$. As can be seen in Fig.~\ref{fig:ecm}(a), the energy smoothly approaches its zero-temperature value and the absence of size-dependent features strongly suggests that there is no phase transition. 
This is also reflected in the specific heat only showing a Schottky-like peak without any finite-size dependence visible on the scale of the plot.
This feature is an expected consequence of a discrete energy spectrum above the ground state.
As the ground state is disordered, in the thermodynamic limit the (staggered) magnetization per site should vanish for all temperatures, and we expect a scaling of $|M_s|/N \sim 1/\sqrt{N} = 1/L$. This is clearly seen in Fig.~\ref{fig:ecm}(b) when plotting $|M_s|/\sqrt{N}$ as a function of $\beta$: All curves for different system sizes practically fall on top of each other, confirming the anticipated scaling. The infinite-temperature ($\beta=0$) value of $|M_s|$ is readily understood using the following simple random walk (RW) argument. A one-dimensional RW taking $N$ random steps $+1$ or $-1$ is, for large $N$, on average a distance of $\sqrt{2N/\pi}$ away from the origin~\footnote{Let the random variable $X_N$ be the position of the RW after $N$ steps. The distribution of $X_N/\sqrt{N}$ approaches a standard normal in the limit $N\to\infty$, and thus for large $N$, $\text{E}[|X_N|/\sqrt{N}]=\sqrt{2/\pi}$, the mean absolute deviation of the standard normal distribution.}, and hence $|M_s|(\beta=0)/\sqrt{N} = \sqrt{2 / \pi}\approx 0.798\dots$
Interestingly, at intermediate temperatures there is a maximum in the staggered magnetization. This indicates that as the temperature is lowered, nearest neighbors locally order antiferromagnetically to some extent, while at even lower temperatures this local order is broken as more diagonals orient antiferromagnetically (see Fig.~\ref{fig:nfmd_and_S}(b) below showing the decline in the fraction of ferromagnetically aligned diagonals as $\beta$ increases), resulting in the decline of $|M_s|$. Note that the location of the maximum in $|M_s|$ is clearly distinct from the one of the specific-heat peak.

\begin{figure}
\includegraphics{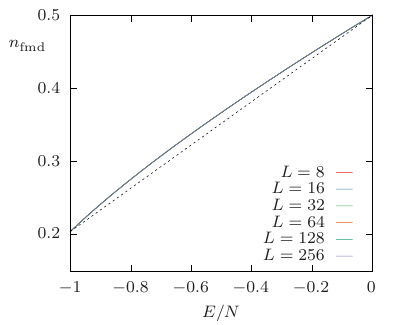}
\caption{Parametric plot showing $\nfmd$ vs.\ $E/N$. The dashed line connects the points $(E/N,\nfmd)$ for infinite and zero temperature, which shows that the relation between the two quantities is close-to-linear. Error bars are smaller than the line width.\label{fig:nfmdVsEnergy}}%
\end{figure}

Next, we turn to the entropy $S$ and the fraction of  diagonal interactions that are ferromagnetically aligned, $\nfmd$.
For both quantities, the authors of Ref.~\cite{Shastry2026} have extracted estimates via the CTMRG method for zero temperature, which we will compare to below. In Fig.~\ref{fig:nfmd_and_S}, we show the results for (a) $S(\beta)/N$ and (b) $\nfmd(\beta)$ for different system sizes $L$. As was the case for the observables discussed above, also these two quantities basically show no system-size dependence when plotted as a function of $\beta$ [cf.\ panels (a) and (b)]. The entropy $S(\beta)/N$ smoothly goes from its infinite-temperature value of $\ln(2)\approx 0.693\,147\dots$ to its zero-temperature value, as seen in panel~(a). The error bars in both cases are not shown as they are smaller than the width of the lines. The estimate for $S(T=0)$ using Eq.~\eqref{eq:pa_S_GS_estimator} (solid line) is in good visual agreement with the value of Ref.~\cite{Shastry2026} (dotted line) even though in the plot we show the case of the relatively small size $L=16$. Similarly, $\nfmd$ is a smooth function going from $1/2$ at $\beta=0$ to its zero-temperature value $\nfmd(T=0)= 0.203\,953\dots$ of Ref.~\cite{Shastry2026} (dotted line) as $\beta\rightarrow\infty$, and it has a similar form to the (total) energy of Fig.~\ref{fig:ecm}(a). To quantify the similarity, we show in Fig.~\ref{fig:nfmdVsEnergy} a parametric plot of $\nfmd$ vs.\ $E/N$. Indeed, the two quantities are almost linearly related. However, the plot clearly deviates from the straight line connecting the infinite- and the zero-temperature points of $(E/N,\nfmd)$, and hence the relation is not perfectly linear.

We extract the ground-state entropy and the zero-temperature value of $\nfmd$ from the final temperature step ($\beta=3$). $S(T=0)$ is calculated using Eq.~\eqref{eq:pa_S_GS_estimator}, whereas $\nfmd(T=0)$ is estimated by averaging over the sampled ground-state configurations (recall that the ground-state energy is known). Zero-temperature results for both quantities are summarized in Table~\ref{tab:resultsEntropy} with the CTMRG results from Ref.~\cite{Shastry2026} quoted below.
For all system sizes $L>32$ the PA data are within one or two error bars of the respective CTMRG value for $L\rightarrow \infty$.
In fact, they do agree to $6-7$ digits.
We also note that already the simple plaquette approximation of the ground states of Ref.~\cite{Meng2008} gives results close to these values for $S(T=0)/N$ [Eq.~\eqref{eq:entropyPlaquetteApprox}] and for $\nfmd$ which is $0.2$ within this approximation.
Note that while not explicitly discussed in Ref.~\cite{Meng2008}, this is easy to derive by noting that two out of the ten allowed plaquette configurations have ferromagnetically aligned diagonals. 

\begin{table}[b]
    \caption{Zero-temperature estimates for the entropy per site $S(T=0)/N$ and the probability for a diagonal to be ferromagnetically aligned $\nfmd(T=0)$ from PAMC simulations at $\beta=3$, the CTMRG results of Ref.~\cite{Shastry2026}, and the plaquette approximation of Ref.~\cite{Meng2008}. $R$ denotes the PA population size, and $M$ the number of independent runs.\label{tab:resultsEntropy}}
    \begin{ruledtabular}
    \begin{tabular}{rllrr}
        \multicolumn{1}{c}{$L$}	& \multicolumn{1}{c}{$S(T=0)/N$} & \multicolumn{1}{c}{$\nfmd(T=0)$} &\multicolumn{1}{c}{$R$} & \multicolumn{1}{c}{$M$} \\\colrule
        $8$	& $0.458\,798\,20(40)$ & $0.204\,114\,1(16)$ & $1.2 \times 10^8$ & $20$ \\
        $16$& $0.458\,777\,13(30)$ & $0.203\,953\,2(17)$ & $4 \times 10^7$ & $20$ \\
        $32$& $0.458\,778\,07(21)$ & $0.203\,952\,7(21)$ & $1 \times 10^7$  & $20$ \\
        $64$& $0.458\,777\,62(12)$ & $0.203\,955\,9(19)$ & $2.5\times 10^6$  & $20$ \\
        $128$& $0.458\,777\,78(10)$ & $0.203\,952\,5(24)$ & $7.5\times 10^5$  & $20$ \\
        $256$& $0.458\,777\,81(12)$ & $0.203\,958\,0(25)$ & $1.85\times 10^5$  & $20$ \\
        $\infty$~\cite{Shastry2026}\!\!\! &$0.458\,777\,72$&$0.203\,953\,87$&&\\	
        $\infty$~\cite{Meng2008}\!\!\! &$0.458\,145\dots$&$0.2$\footnote{Reference~\cite{Meng2008} does not consider $\nfmd$, but as 2 of the 10 allowed plaquette configurations feature ferromagnetically oriented diagonals within their approximation $\nfmd = 1/5$.}&&
    \end{tabular}
    \end{ruledtabular}
\end{table}

\begin{figure*}[ht]
    \includegraphics{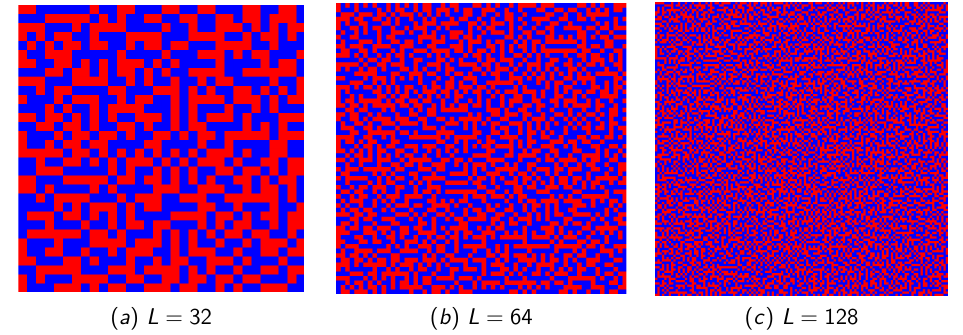}
    \caption{Exemplary ground-state configurations for $L=32$, $64$, and $128$.\label{fig:groundstateConfigs}}
\end{figure*}
For a better understanding of the nature of the ground states, we now turn to studying the sampled spin configurations with minimal energy -- noting that these are not accessible by the methods employed in Ref.~\cite{Shastry2026}. Exemplary zero-temperature spin configurations of different linear sizes $L$ are presented in Fig.~\ref{fig:groundstateConfigs}. These clearly show no long-range order, and appear increasingly random as the system size increases. 
One disadvantage of directly considering the spin configurations is that it is not immediately visible which diagonal spins interact and which do not.

\begin{figure}
    \includegraphics{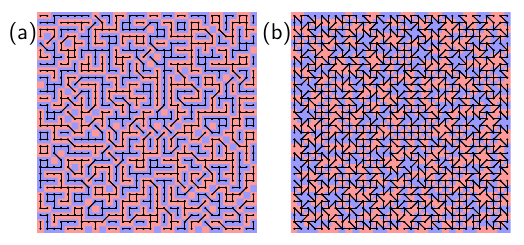} %
    \caption{(a) Frustrated and (b) satisfied interactions of the $L=32$ ground-state configuration of Fig.~\ref{fig:groundstateConfigs}(a) indicated with black lines. For reference the spin configuration is shown again in lighter colors.\label{fig:groundstateConfigsBonds}}
\end{figure}
To better distinguish interacting from non-interacting diagonals, we show in Fig.~\ref{fig:groundstateConfigsBonds}(a) the frustrated interactions and (b) the satisfied interactions overlaid on top of the $L=32$ spin configuration of Fig.~\ref{fig:groundstateConfigs}(a). Also from the bond configurations the absence of conventional long-range order can be seen quite clearly. Similar to the Ising antiferromagnet on the triangular lattice, the frustrated interactions lead to a labyrinth pattern. 
However, unlike the triangular lattice, the connected segments of frustrated bonds here form only small to mid-sized networks that do not span the system.
This is in agreement with the report of a short correlation length in the ground state of the present model~\cite{Shastry2026}, whereas the ground state of the Ising antiferromagnet on the triangular lattice shows a diverging correlation length associated with critical behavior~\cite{Bloete1993}. The counterpart to the labyrinth pattern of frustrated interactions is the densely connected network of satisfied interactions shown in panel~(b). In this representation one easily recognizes some reoccurring local structures, such as checkerboard patterns, which are two or three lattice spacings wide in one direction and extend over a longer distance in the other. In the bond representation in these regions all nearest-neighbor interactions are satisfied. Another such pattern is a ferromagnetically oriented $2\times 2$ square with spins on its diagonals pointing in the opposite direction, resulting in a ``rotating star'' local bond configuration.

\begin{figure}[ht]
    \includegraphics{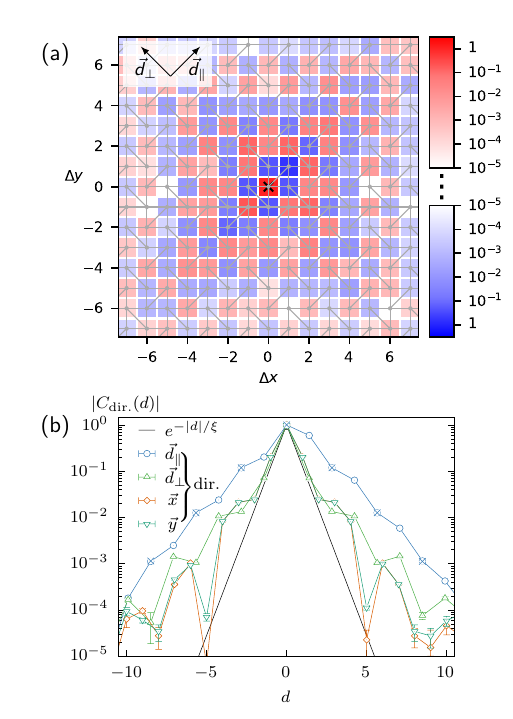}
    \caption{Spin-spin correlation for $L=64$ for spins of one sublattice with other spins separated by the distance vector $\vec d = (\Delta x,\Delta y)$. (a) Spatial correlation matrix, in which the sign of the correlation is indicated by a red (blue) color for positive (negative) correlation and its absolute value determines the color intensity. The (antiferromagnetic) diagonal interaction corresponds to the $\vec d=(1,1)$ entry (see lattice structure indicated by the gray overlay and the inset showing the axes parallel and perpendicular to the $(1,1)$ direction). The origin is marked by a cross ($\times$). (b) The absolute value of the correlation along the two diagonals and the two lattice directions (see text for details). The (blue) crosses show $C_{\vec{d}_\parallel}(-d)$ for even $d$, illustrating the $\pm d $ symmetry of the correlation function along the interacting diagonal direction. The solid black line indicates a purely exponential decay with $\xi\approx0.476\,859$~\cite{Shastry2026}.\label{fig:correlation}}
\end{figure}

For more quantitative evidence of the short-range nature of the ground-state correlations, we proceed by studying the spatial spin-spin correlation function,
\begin{equation}
C(\Delta x, \Delta y) = \langle \sigma_{\vec{r}(i)} \sigma_{{\vec{r}(i)}+(\Delta x, \Delta y)} \rangle_A \label{eq:correlationFct},
\end{equation}
where $\sigma_{\vec{r}(i)}$ represents the spin at the two-dimensional lattice position vector of site $i$~\footnote{Note that $\sigma_{i}$ is to be understood as a shorthand notation of $\sigma_{\vec{r}(i)}$ and the two denote the same object. Also note that the $+$ operator in Eq.~\eqref{eq:correlationFct} is defined such that it accounts for PBC}. $\langle \cdot \rangle_A$ denotes ensemble averaging over sites $i$ restricted to sublattice A of the four sublattices depicted in Fig.~\ref{fig:latticeDomainDecomposition} with this sublattice containing all sites with a diagonal interaction in the (positive) (1,1) direction (blue circles in Fig.~\ref{fig:latticeDomainDecomposition}). (This choice is such that the averaging is performed over equivalent lattice sites, and is not related to the GPU implementation.) Restricting $\sigma_{\vec{r}(i)}$ to one sublattice is necessary as otherwise it would not be possible to distinguish correlations in the direction of the interacting diagonal and in the one perpendicular to it. One should note that sublattice~A was chosen without loss of generality, and any other sublattice gives equivalent results with the correlation matrix trivially being rotated by multiples of $\pi/2$. Note that while $\sigma_{\vec{r}(i)}$ is restricted to one sublattice, the same is not necessary for $(\Delta x,\Delta y)$ in $\sigma_{\vec{r}(i)+(\Delta x,\Delta y)}$, which in principle runs over the full lattice. For practical reasons, we have only calculated $C(\Delta x,\Delta y)$ for $\Delta x, \Delta y \in \{-15,-14,\dots,0,\dots,14,15\}$.

The results of this evaluation are depicted in Fig.~\ref{fig:correlation}. Panel~(a) shows the values of the correlation matrix as a heat map with different shades of red (blue) indicating positive (negative) correlation. Lighter colors correspond to a smaller absolute value of the correlation. When the obtained error bars exceed one third of the absolute value of the correlation, the value was marked as a white square as based on the numerical data the sign cannot be determined beyond reasonable doubt. The lattice structure is indicated by gray lines, and the origin is marked by a cross. While the correlation changes between positive and negative correlation when moving away from the origin, it does not seem to follow a simple pattern in general. Only along the direction of the interacting diagonal, the data are consistent with a sign change at every other step, specifically whenever passing a diagonal interaction.
For symmetry reasons, it is easy to see that the correlation matrix should be symmetric with respect to the main diagonal, the direction of the diagonal interaction, i.e., $\vec{d}_\parallel = (1,1)$. The data are in good agreement apart from some deviations further away from the origin due to the limited numerical accuracy.

Figure~\ref{fig:correlation}(b) shows the absolute value of the measured spin-spin correlation as a function of distance $d$ along the two diagonal directions, the main diagonal $\vec{d}_\parallel$ and the one perpendicular to it, $\vec{d}_\perp$, and the two lattice directions $\vec{x}$ and $\vec{y}$. The solid line shows $e^{-|d|/\xi}$ for $\xi\approx 0.476\,859$~\cite{Shastry2026}. First, note that irrespective of the direction, the correlation is consistent with an at least exponential decay, in line with the previous observations suggesting only short-range correlations. As a consequence of the symmetry with respect to the main diagonal, the functions for the $x$- and $y$-directions should be equivalent, and the correlation function along $\vec{d}_\perp$ has to be symmetric with respect to the origin, i.e., $C_{\vec{x}}(d)=C_{\vec{y}}(d)$ and $C_{\vec{d}_\perp}(d)=C_{\vec{d}_\perp}(-d)$. Both of these (again up to some numerical deviations further away from the origin) are clearly seen in the data for the $|C_\text{dir.}(d)|$ functions. Additionally, we observe that also the functions in the $x$- and $y$-directions are symmetric with respect to $d=0$.
Surprisingly, even along the main diagonal we numerically observe $C_{\vec{d}_\parallel}(d)=C_{\vec{d}_\parallel}(-d)$ for every other point (see the crosses ($\times$) showing $C_{\vec{d}_\parallel}(-d)$ for even $d$). Furthermore, we note that $|C_{\vec{d}_\parallel}(d)|$ alternates between smaller and larger `steps' with smaller ones occurring when passing over an interacting diagonal and larger ones seen when passing over a diagonal without an interaction.
For small $d$, $|C_\text{dir.}(d)|$ is somewhat consistent with an exponential decay with a correlation length as calculated in Ref.~\cite{Shastry2026}. We note that evaluating the decay of the correlation function is somewhat difficult due to the irregular sign changes.
An artifact of this can be seen for $d=5$ in the two lattice directions for which the absolute value of the correlation function is particularly small.  
This is different for the main diagonal direction for which we observe more regular sign changes, and hence $|C_\text{dir.}(d)|$ more closely resembles a straight line, albeit suggesting a larger correlation length $\xi \approx 1$.

\begin{figure}
    \includegraphics{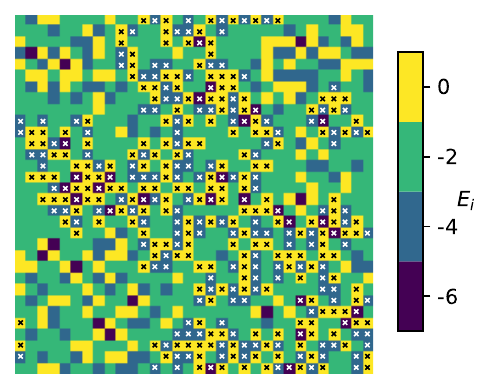}
    \caption{Local-energy configuration of the $L=32$ ground state from Fig.~\ref{fig:groundstateConfigs}(a). The sites marked with a cross ($\times$) denote a randomly chosen cluster of sites with $E_i\neq -2$.\label{fig:groundstateConfigsEnergy}}
\end{figure}

Last, we consider the sampled ground-state configurations in a `local'-energy representation $E_i=\sigma_i \times (\sum_{j \in \nn} \sigma_j + 2\alpha \sigma_k)$ summing over the four \nn{}s $\sigma_j$ and the diagonally connected $\sigma_k$.
As $\alpha=1$, $E_i$ only can take values $-6,-4,-2,0,2,4$ and $6$. Note that when summing over all `local' energies, each interaction is counted twice. Hence, the ground-state energy of $-1$ per site corresponds to a local energy of $E_i=-2$. When inspecting the local energies of ground states (see Fig.~\ref{fig:groundstateConfigsEnergy} showing an exemplary local-energy configuration for $L=32$), one finds that all $E_i \leq 0$ can occur. The local $E_i$ is $-6$ when all interacting spins point in the opposite direction, $-4$ when one of the \nn{} interactions is frustrated, $-2$ when either two \nn{} interactions are frustrated or the diagonal one, and $0$ when either three \nn{} are frustrated or the diagonal and one \nn{} interaction. $E_i>0$ does not occur in any of the ground states.

\begin{table}[b]
    \caption{Estimate for the frequencies $P(E_i)$ at which local energies $E_i \in \{-6,-4,-2,0\}$ in ground states are observed.~\label{tab:resultsEnergyFreq}}
    \begin{ruledtabular}
    \begin{tabular}{rllll}
        \multicolumn{1}{c}{$L$}	& \multicolumn{1}{c}{$p(E_i=-6)$} & \multicolumn{1}{c}{$p(E_i=-4)$} &\multicolumn{1}{c}{$p(E_i=-2)$} & \multicolumn{1}{c}{$p(E_i=0)$} \\\colrule
        $32$ & $0.0421382(7)$ & $0.2368569(12)$ & $0.3998718(12)$ & $0.3211332(7)$\\
        $64$ & $0.0421383(7) $ & $0.2368561(12)$ & $0.3998730(17)$ & $0.3211326(10)$\\
        \!\!$128$ & $0.0421380(8)$ & $0.2368558(12)$ & $0.3998744(21)$ & $0.3211318(13)$\\
        \!\!$256$ & $0.0421384(7)$ & $0.2368554(12)$ & $0.3998739(23)$ & $0.3211323(13)$\\
    \end{tabular}
    \end{ruledtabular}
\end{table}

\begin{figure}[tb]
    \includegraphics{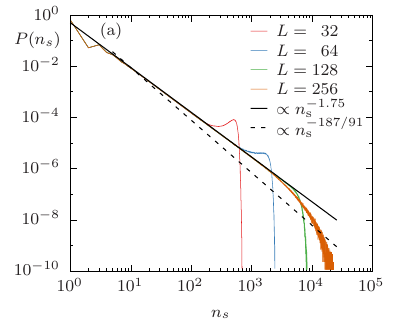}
    \includegraphics{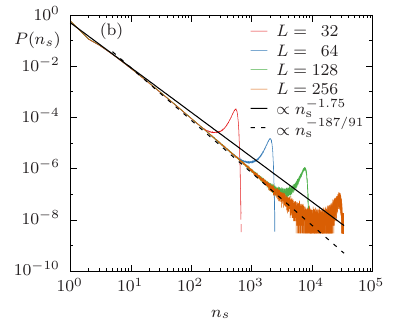}
    \caption{Cluster-size distribution of (a) clusters of sites with $E_i\neq-2$ (such as the one shown in Fig.~\ref{fig:groundstateConfigsEnergy}) for different system sizes, and (b) (uncorrelated) random clusters on the square lattice at the site-percolation threshold $p_c\approx 0.5927\dots$. In both panels the expected asymptotic scaling for ordinary percolation in two dimensions of $\propto n_s^{-187/91}$ is marked by a dashed line, and the empirically observed scaling of $P(n_s)\propto n_s^{-1.75}$ by a solid line.\label{fig:clusterSize}}
\end{figure}

As in the other representations, the configuration shows no signs of conventional long-range order. We have estimated the frequency at which each energy occurs (see Table~\ref{tab:resultsEnergyFreq}). Just under 40\% of sites have a local energy of $E_i=-2$ (the mean (local) energy in the ground state). We have carried out a cluster analysis of sites connected (by the square lattice structure) that share the same value of $E_i$. The size of such a cluster refers to the number of sites $n_s$ contained within. With the exception of $E_i=-6$ only occurring at isolated sites, i.e., $n_s=1$, one finds the cluster-size distribution to decay exponentially fast (see Appendix~\ref{app:clusterSizeIndEnergies}). Finally, we consider clusters of sites with $E_i \neq -2$ (all sites with local energies different from the ground-state energy). Here, we find good numerical resemblance with a power-law decay $P(n_s)\sim n_s^{-\tau}$ with an exponent $\tau=1.75$, cf.\ Fig.~\ref{fig:clusterSize}(a). Finding a power-law distribution in this case is not surprising per se, as $p(E_i \neq -2) = 1-p(E_i=-2) \approx 0.6001\dots$ is quite close to the ordinary site-percolation threshold of the square lattice of $0.5927\dots$~\cite{Newman2001}. However, the observed exponent does not agree at all with the respective exponent from ordinary percolation $\tau=187/91\approx 2.05$, clearly indicating that the effect is not related to ordinary percolation. For comparison, we have sampled random clusters for the same system sizes, the results of which are presented in Fig.~\ref{fig:clusterSize}(b). This clearly shows that the scaling in the two cases is distinctly different. We are not aware of any related model with a similar cluster-size exponent, and we cannot comment on the origin of the observed exponent. While it is in principle possible that this observation is a finite-size effect and that asymptotically the distribution becomes exponential, such a scenario would be challenging to observe numerically due to the proximity of $p(E_i\neq -2)$ to the site-percolation threshold.

\section{Conclusion and outlook}\label{sec:conclusionAndOutlook}

We have demonstrated how the population annealing method can be used to sample ground states of complex systems in statistical mechanics, and how to estimate the ground-state entropy. This approach has been applied to the Ising antiferromagnet on the Shastry-Sutherland lattice for $\alpha=1$, which is known to have a non-zero entropy per site at zero temperature. We find that our numerical results are in excellent agreement with a recent high-precision CTMRG calculation~\cite{Shastry2026} demonstrating the efficacy of this approach as a general method to calculate ground-state properties including the residual entropy. Another application of this method is the Ising model on the honeycomb lattice with frustrated first- and second-neighbor interactions, for which we also estimate the ground-state entropy, see Ref.~\cite{Gessert2026}.

In contrast to the methods used in previous studies of this model, population annealing allows us to sample concrete ground-state configurations, which we studied in this work. While most results are consistent with the picture of thermodynamically trivial behavior with a short-range-correlated and disordered ground state, we found a non-trivial scaling in the size distribution of clusters in the local-energy representation: The size distribution of clusters of sites that have a local energy $E_i$ different from the ground-state energy (up to the trivial double-counting factor 2) is consistent with a power-law decay with an exponent $\tau=1.75$. This value is distinctly different from the ordinary percolation exponent. Further work is required to understand the origin and the genuity of this power-law scaling.

We also note that surprisingly little is known about the thermal transition towards the Néel state for $\alpha<1$. Clearly, for $\alpha=0$ Eq.~\eqref{eq:Hamiltonian} is identical to the Ising AFM on the square lattice that has a phase transition at $T_c=2/\ln(1+\sqrt{2})=2.26919\dots$ with critical exponents in the Ising universality class. As $\alpha$ increases, the transition temperature is expected to decrease, and it is expected to finally reach zero temperature for $\alpha=1$. However, whether the transition stays within the Ising universality class remains unexplored and would be of interest for a future study. The closely related Ising AFM on the square lattice with second-neighbor interactions is believed to have a tricritical point on the Néel transition line~\cite{Jin2012}, which makes this a plausible scenario here, too. If there was such a tricritical point $\alpha^\ast$, then one expects that $\alpha^\ast \in (0.5,1)$. This is due to the fact that $\alpha=0.5$, the special case for which all interactions have the same strength, was studied previously in Ref.~\cite{Yu2015} indicating that the transition at $T_c=1.261(1)$ is within the Ising universality class. In fact, the Ising FM on the same lattice has been studied, and its critical temperature has been calculated exactly, i.e., $T_c = 2.9263\dots$~\cite{Codello2010}. The equation underlying this result also has a solution for negative coupling ratios (not discussed in Ref.~\cite{Codello2010}). This was recently also noted by the authors of Ref.~\cite{Zinati2026} (including the author of Ref.~\cite{Codello2010}), resulting in a transition temperature $T_c=1.26194\dots$, which is in very good agreement with the numerical estimate of Ref.~\cite{Yu2015}. We note that in Refs.~\cite{Yu2015,Codello2010,Zinati2026} this model was studied in the context of Archimedean lattices, with the Shastry-Sutherland lattice being one of the eleven Archimedean lattices. The lattice is denoted by the label T3 in Ref.~\cite{Yu2015}, and by $(3^2,4,3,4)$ in Refs.~\cite{Codello2010,Zinati2026}.

~

\begin{acknowledgments}
    W.J. was supported by the Deutsch-Franz\"osische Hochschule (DFH-UFA) under Grant No.\ CDFA-02-07. D.G. and M.W. thank the DAAD (Project ID 57807776) and SPARC (SPARC-GIANT/2025-2026/PG250011) for support through an Indo-German grant scheme.
\end{acknowledgments}

~
\appendix
\section{Cluster-size distribution for individual local energy values}\label{app:clusterSizeIndEnergies}
\begin{figure*}
    \includegraphics{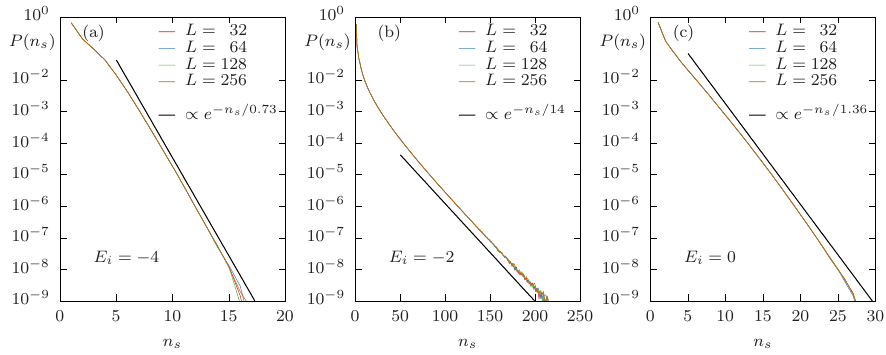}
    \caption{Cluster-size distribution of the local-energy configurations for (a) $E_i=-4$, (b) $E_i=-2$, and (c) $E_i=0$ and for different sizes $L$. In each panel the solid black line is a guide to the eye indicating a purely exponential decay.\label{fig:clusterSize_energy_ind}}
\end{figure*}

Figure~\ref{fig:clusterSize_energy_ind} shows the cluster-size distribution determined from the local-energy configurations for the different values of $E_i$. Note that the distribution for $E_i=-6$ is trivial, as in this case always $n_s=1$. Unlike the distribution for $E_i \neq -2$ shown in Fig.~\ref{fig:clusterSize}(a), these distributions decay at least exponentially fast (solid black lines).

\end{document}